\documentclass{amsart}

\usepackage{fullpage}
\usepackage{amsmath,amsthm,amssymb}
\usepackage{hyperref}

\usepackage[T5,T1]{fontenc}
\usepackage[utf8]{inputenc}
\usepackage{lmodern}

\usepackage{tikz-cd}
\usepackage{proof}
\usepackage{mathtools}
\usepackage{csquotes}

\theoremstyle{plain}
\newtheorem{theorem}{Theorem}[section]

\theoremstyle{definition}

\title{On the complexity of normalization for the planar $\lambda$-calculus}
\author{Anupam Das, Damiano Mazza, {\fontencoding{T5}\selectfont Lê Thành Dũng 
(Tito) Nguyễn}, Noam Zeilberger}

\begin{document}

\maketitle

Recall that an untyped $\lambda$-term $t$ is \emph{linear} if there exists a list $\Gamma$ -- the list of free variables in $t$ -- such that $\Gamma\vdash t$ is derivable with the rules below (with $\Gamma$ and $\Delta$ disjoint in $app$):
\[
  \infer[var]{x \vdash x}{}
  \qquad
  \infer[app]{\Gamma,\Delta \vdash t\,u}{\Gamma \vdash t & \Delta \vdash u}
  \qquad
  \infer[lam]{\Gamma \vdash \lambda x.t}{\Gamma,x \vdash t}
  \qquad
  \infer[exc]{\Gamma,x,y,\Delta \vdash t}{\Gamma,y,x,\Delta \vdash t}
\]
Call a linear $\lambda$-term $t$ \emph{planar} when there is an \emph{ordered} list $\Gamma$ such that $\Gamma\vdash t$ is derivable in the subsystem \emph{without the $exc$ rule}: for example, $\lambda x.\, \lambda y.\, f\,x\,y$ is planar but $\lambda x.\, \lambda y.\, f\,y\,x$ is not. Planar $\lambda$-terms are closed under $\beta$-reduction. Furthermore, this notion is motivated by semantics (non-symmetric monoidal closed categories), topology (a linear $\lambda$-term is planar when its representation as a syntax tree with binding edges is a planar combinatorial map) and linguistics (in the Lambek calculus~\cite{Lambek}, a precursor of linear logic).

Less attention has been paid, however, to the \emph{computational} consequences of planarity. There is a recent implicit complexity result~\cite{aperiodic} using planar $\lambda$-terms, where general linear $\lambda$-terms would be too expressive. Here, we focus on the \emph{complexity} of \emph{normalizing} $\lambda$-terms, asking ourselves whether planarity lowers it. For linear (possibly non-planar) $\lambda$-terms, we know that:
\begin{theorem}[{\cite{Mairson2004}}]\label{thm:mairson}
  The following decision problem is $\mathsf{P}$-complete under logarithmic space reductions:
  \begin{itemize}
    \item Input: two (untyped) linear $\lambda$-terms $t$ and $u$.
    \item Output: are $t$ and $u$ $\beta$-convertible, that is, do they have the same normal form?
  \end{itemize}
\end{theorem}
(Note that the complexity of the $\beta$-convertibility problem for simply typed (possibly non-linear) $\lambda$-terms is much higher, namely $\mathsf{TOWER}$-complete -- this is implicit in~\cite{Statman1979}, as explained in~\cite{SafeComplexity}.)

\textbf{We believe that this problem is still $\mathsf{P}$-complete when $t$ and $u$ are planar.} Two years ago, we claimed this as a theorem\footnote{In a talk at the Structure Meets Power 2021 workshop: \url{http://noamz.org/talks/smp.2021.06.28.pdf}} but the proposed proof -- which purported to provide a logspace reduction from the Circuit Value Problem (CVP), just like Mairson's proof of Theorem~\ref{thm:mairson} -- contained a subtle yet serious flaw, described at the end of Section~\ref{sec:bool}.

In this extended abstract, we outline another attempt to reduce CVP to planar normalization.

\section{The Circuit Value Problem}\label{sec:cvp}

For our purposes, a \emph{boolean circuit} with $n$ gates can be seen as a list of $n$ equations defining the values of the boolean variables $x_1,\dots,x_n$, such as the following example:
\[ x_1 := 1;\; x_2 := 0;\; x_3 := 1;\; x_4 = x_1 \land x_2;\; x_5 = \lnot x_1;\; x_6 = x_5 \land x_3;\; x_7 = x_4 \lor x_6 \]
Here, equations 4 to 7 define $x_7 = (x_1 \land x_2) \lor (\lnot x_1 \land x_3) = \mathsf{if}\; x_1\; \mathsf{then}\; x_2\;\mathsf{else}\; x_3$, so the final result of the circuit is $(\mathsf{if}\; 1\; \mathsf{then}\; 0\;\mathsf{else}\; 1) = 0$.
In each equation, the right-hand side contains either a constant $0/1$ or the application of an operator $\lnot,\land,\lor$. Furthermore, we require that in the latter case, the arguments given to the operator have been defined \emph{before} the current equation; in other words, the enumeration $x_1, x_2, \dots$ is a \emph{topological ordering} of the circuit.
\begin{theorem}[{\cite[Theorem~6.2.1]{LimitsParallel}}]
  The \emph{Topologically Ordered Circuit Value Problem (TopCVP)}, defined below, is $\mathsf{P}$-complete.
  \begin{itemize}
    \item Input: a topologically ordered boolean circuit, as in the above example.
    \item Output: the final value computed by the circuit.
  \end{itemize}
\end{theorem}

\section{Planar booleans do not suffice}\label{sec:bool}

To encode the Circuit Value Problem in the linear $\lambda$-calculus, Mairson~\cite{Mairson2004} uses a linear encoding of booleans. Unfortunately, his encoding represents 0 as a non-planar $\lambda$-term, namely $\lambda x.\, \lambda y.\, \lambda f.\, f\,y\,x$.

A planar linear encoding of booleans has been introduced in~\cite[Chapter~7]{titoPhD} to give a strictly \emph{linear} variant of the previously mentioned result of~\cite{aperiodic}, whose original statement used planar \emph{affine} $\lambda$-terms.
\[ \mathtt{false} = \lambda k.\, \lambda f.\, k\, f\, (\lambda x.\, x) \qquad \mathtt{true} = \lambda k.\, \lambda f.\, k\, (\lambda x.\, x)\, f \]
While our reduction targets untyped $\lambda$-terms, it can be useful to think of these terms as the only inhabitants in normal form of the type
\[ \mathtt{Bool} = \forall\alpha\beta.\,((\alpha\multimap\alpha)\multimap(\alpha\multimap\alpha)\multimap\beta)\multimap(\alpha\multimap\alpha)\multimap\beta \]
This can be seen as the image, by a continuation-passing-style transformation, of an encoding using linear $\lambda$-terms \emph{with pairs} proposed by Matsuoka~\cite{Matsuoka2015} in his alternative proof of Theorem~\ref{thm:mairson}:
\[ \mathtt{false}' = \lambda f.\, (f,\, \lambda x.\, x) \qquad \mathtt{true}' = \lambda f.\, (\lambda x.\, x,\, f) \qquad \mathtt{Bool}' = \forall\alpha.\, (\alpha\multimap\alpha)\multimap(\alpha\multimap\alpha)\otimes(\alpha\multimap\alpha)\]
We can also define boolean connectives acting on the encodings of~\cite{titoPhD} (we have $\mathtt{cstt}\,b =_\beta \mathtt{true}$ and $\mathtt{cstf}\,b =_\beta \mathtt{false}$ for $b\in\{\mathtt{true,false}\}$), using the notations $\mathtt{id} = \lambda x.\, x$ and $f \circ g = \lambda x.\, f\, (g\, x)$:
\begin{align*}
\mathtt{cstt} &= \lambda b.\, \lambda k.\, \lambda f.\, b\, (\lambda g.\, \lambda h.\, k\, \mathtt{id}\, (g \circ h))\, f\\
\mathtt{cstf} &= \lambda b.\, \lambda k.\, \lambda f.\, b\, (\lambda g.\, \lambda h.\, k\, (g \circ h)\, \mathtt{id})\, f\\
\mathtt{not} &= \lambda b.\, b\,(\lambda g.\, \lambda h.\, g\, (\mathtt{cstt}\, (h\, \mathtt{true})))\, \mathtt{cstf}\\
\mathtt{and} &= \lambda b_1.\, \lambda b_2.\, \lambda k.\, b_1\, (\lambda f_1.\, b_2\, (\lambda f_2.\, \lambda f_3.\, k\, (\lambda x.\, f_1\, (f_2\, x))\, f_3))
\end{align*}
(disjunction can be derived by De Morgan's laws). This is enough to translate boolean \emph{formulas} into planar linear $\lambda$-terms.

However, to transpose Mairson's methodology for encoding boolean circuits to the planar linear setting, we would need a planar $\lambda$-term $\mathtt{copy}$ such that (similarly to the $\mathsf{W}$ combinator in Curry's $\mathsf{BCKW}$)
\[ \forall t\in\{\mathtt{true,false}\},\; \mathtt{copy}\,f\,t =_\beta f\,t\,t \]
We have not been able to find such a term; we did manage to define a planar $\lambda$-term $\mathtt{copy}'$ that satisfies $\mathtt{copy}'\,t\,f =_\beta f\,t\,t$, but this is significantly different in a planar setting. Hence the gap in our previous attempt at reducing CVP to $\beta$-convertibility of planar $\lambda$-terms.

\section{A new encoding of TopCVP}

Our new idea is to work with an encoding of \emph{bit vectors}, on which we implement the following operations:
  \[ \mathsf{not}_{i,n}(\langle x_1,\dots,x_n \rangle) = \langle x_1,\,\dots,\,x_n,\,\lnot x_i \rangle \qquad
   \mathsf{and}_{i,j,n}(\langle x_1,\dots,x_n \rangle) = \langle x_1,\,\dots,\,x_n,\,x_i \land x_j \rangle \]
and the analogous $\mathsf{or}_{i,j,n}, \mathsf{false}_n, \mathsf{true}_n \colon \{0,1\}^n \to \{0,1\}^{n+1}$ for $1 \leq i,j \leq n$. Let also $\mathsf{last}_n(\langle x_1,\dots,x_n\rangle) = x_n$.

The value of our example circuit from Section~\ref{sec:cvp} can then be expressed, using these operations, as
\[ \mathsf{last}_7 \circ \mathsf{or}_{4,6,6} \circ \mathsf{and}_{5,3,5} \circ \mathsf{not}_{1,4} \circ \mathsf{and}_{1,2,3} \circ \mathsf{true}_2 \circ \mathsf{false}_1 \circ \mathsf{true}_0 (\langle\rangle) \]

\subsection{Representation of bit vectors}

Unsurprisingly, we use the Church encoding of $k$-tuples, together with the above-mentioned type $\mathtt{Bool}$, to represent vectors of $k$ bits. For instance, $\langle 0,1,0 \rangle$ is encoded as
\[ \overline{\langle 0,1,0 \rangle} = \lambda k.\, k\, \mathtt{false}\, \mathtt{true}\, \mathtt{false} \]
which should be seen as an inhabitant of the type
\begin{align*}
  \mathtt{Bool}_3 &= \forall\gamma.\, (\mathtt{Bool} \multimap \mathtt{Bool} \multimap \mathtt{Bool} \multimap \gamma) \multimap \gamma
\end{align*}

\subsection{Implementing vectorial operations}

First, we implement $\mathsf{fetch}_{i,n}(\langle x_1,\dots,x_n\rangle) = \langle x_1, \dots, x_n, x_i \rangle$:
\begin{align*}
  \mathtt{fetch}_{i,n} &= \lambda v.\, v\, (\lambda x_1.\, \dots\, \lambda x_n.\, x_1\, c_1\, f_1\, (\dots (x_n\, c_n\, f_n\, \overline{\langle 0,\dots,0 \rangle} )\dots))\\
  \text{where}\ c_j &= \begin{cases}
    \lambda g.\, \lambda h.\, (g \circ (\lambda k.\,\lambda b_1.\, \dots\, \lambda b_{n+1}.\, k\, b_1 \dots b_{i-1}\, (\mathtt{cstt}\, b_i)\, b_{i+1}\,\dots\, b_{n}\, (\mathtt{cstt}\, b_{n+1})) \circ h) & \text{if}\ i = j\\
    \lambda g.\, \lambda h.\, (g \circ (\lambda k.\,\lambda b_1.\, \dots\, \lambda b_{n+1}.\, k\, b_1 \dots b_{j-1}\, (\mathtt{cstt}\, b_j)\, b_{j+1}\,\dots\, b_{n+1}) \circ h) & \text{otherwise}
  \end{cases}\\
  \text{and}\ f_j &= {\begin{cases}
    \lambda k.\,\lambda b_1.\, \dots\, \lambda b_{n+1}.\, k\, b_1 \dots b_{i-1}\, (\mathtt{cstf}\, b_i)\, b_{i+1}\,\dots\, b_{n}\, (\mathtt{cstf}\, b_{n+1}) & \text{if}\ i = j\\
    \lambda k.\,\lambda b_1.\, \dots\, \lambda b_{n+1}.\, k\, b_1 \dots b_{j-1}\, (\mathtt{cstf}\, b_j)\, b_{j+1}\,\dots\, b_{n+1} & \text{otherwise}  
  \end{cases}}
\end{align*}
Note that by replacing every $\mathtt{cstt}\, b_{n+1}$ by $\mathtt{cstf}\, b_{n+1}$ and vice versa, we get an implementation of $\mathsf{not}_{i,n}$!

We then set $\mathtt{and}_{i,j,n} = \mathtt{and}'_n \circ \mathtt{fetch}_{i,n+1} \circ \mathtt{fetch}_{j,n}$ where $\mathtt{and}'_n$ implements an in-place conjunction
\[ \mathsf{and}'_n(\langle x_1, \dots, x_{n+2}\rangle) = \langle x_1,\, \dots,\, x_n,\, x_{n+1} \land x_{n+2} \rangle \]
To define $\mathtt{and}'_n$, we reuse the planar $\lambda$-term $\mathtt{and}$ that implements the conjunction on the booleans of Section~\ref{sec:bool}:
\[ \mathtt{and}'_n = \lambda v.\, \lambda k.\, v\, (\lambda x_1.\, \dots\, \lambda x_{n+2}.\, k\, x_1\, \dots\, x_n\, (\mathtt{and}\, x_{n+1}\, x_{n+2})) \]
Finally, we take:
\[ \mathtt{last}_{i,n} = \lambda v.\, v\, (\lambda x_1.\, \dots\, \lambda x_n.\, x_1\, (\lambda g.\, \lambda h.\, g \circ h)\, \mathtt{id}\, (\dots (x_{n-1}\, (\lambda g.\, \lambda h.\, g \circ h)\, \mathtt{id}\, x_n)\dots)) \]

\bibliographystyle{alphaurl}
\bibliography{refs.bib}

\end{document}